\documentclass{pasj01}
\usepackage[dvipdfmx]{}
\usepackage{natbib,bm,url,color}
\usepackage{ulem}
\Received{$\langle$reception date$\rangle$}
\Accepted{$\langle$acception date$\rangle$}
\Published{$\langle$publication date$\rangle$}

\begin{document}

\title{Gaia's Detectability of Black Hole-Main Sequence Star Binaries Formed in Open Clusters}
\author{Minori Shikauchi\altaffilmark{1,2*}, Jun Kumamoto\altaffilmark{3}, Ataru Tanikawa\altaffilmark{4,5}, Michiko S. Fujii\altaffilmark{3}}
\altaffiltext{1}{Department of Physics, the University of Tokyo, 
7-3-1 Hongo, Bunkyo, Tokyo 113-0033, Japan}
\altaffiltext{2}{Research Center for the Early Universe, the University of Tokyo, 
7-3-1 Hongo, Bunkyo, Tokyo 113-0033, Japan}
\altaffiltext{3}{Department of Astronomy, the University of Tokyo, 
7-3-1 Hongo, Bunkyo, Tokyo 113-0033, Japan}
\altaffiltext{4}{Department of Earth Science and Astronomy, College of Arts and Sciences, the University of Tokyo, 
3-8-1 Komaba, Meguro, Tokyo 153-8902, Japan}
\altaffiltext{5}{RIKEN Center for Computational Science, 7-1-26 Minatojima-Minami-machi, Chuo, Kobe, Hyogo 650-0047, Japan}
\email{shikauchi@resceu.s.u-tokyo.ac.jp}


\maketitle

\begin{abstract}
Black hole-main sequence star (BH-MS) binaries are one of the targets of the future data releases of the astrometric satellite {\it Gaia}. They are supposed to be formed in two main sites: a galactic field and star clusters. However, previous work has never predicted the number of BH-MS binaries originating in the latter site. In this paper, we estimate the number of BH-MS binaries formed in open clusters and detectable with {\it Gaia} based on the results of {\it N}-body simulations. By considering interstellar extinction in the Milky Way (MW) and observational constraints, we predict $\sim 10$ BH-MS binaries are observable. 
We also find that chemical abundance patterns of companion MSs will help us to identify the origin of the binaries as star clusters. Such MSs are not polluted by outflows of the BH progenitors, such as stellar winds and supernova ejecta. Chemical anomalies might be a good test to confirm the origin of binaries with relatively less massive MSs ($\lesssim 5M_{\odot}$), orbital periods ($\sim 1.5\;$year) and higher eccentricities ($e \gtrsim 0.1$).
\end{abstract}

\section{Introduction}
Stellar mass black holes (BHs) are the remnants of massive stars. 
They have been detected as X-ray binaries in the Milky Way (MW), which have very short period such as several hours. 
The number of discovered BHs in this way is less than 100 \citep{BlackCAT2016}, while theoretical studies have estimated the total number of stellar mass BHs in the MW to be $10^8$ -- $10^9$ \citep{Shapiro1983,vandenHeuvel1992,BrownandBethe1994,Samland1998,Agoletal2002}.

On the other hand, those in extragalactic distances have been detected by gravitational wave observations \citep{GWTC2019}. These binary BHs also have very short periods and therefore merge by emitting gravitational waves. 
Thus, observed BHs are so far biased to those with short periods.

There are the other two ways to find binaries including BHs (hereafter, BH
binaries) especially with longer orbital periods. 
One is radial velocity observations. There are already a few detection reports of long orbital period BH binaries with this method  \citep{Giesers2018,Thompson2019,Liu2019}. In particular, \cite{Liu2019} have reported a $70M_\odot$ BH in a binary system \citep[but see][]{ Eldridge2019,tanikawa2019,Safarzadeh2019,El-Badry2020,Irrgang2020}.
The other is astrometric observations. \cite{GouldandSalim2002} first investigated black holes not producing supernovae by using data of Hipparcos and showed the successor observation will observe their companion stars.
{\it Gaia} mission, the successor of Hipparcos, \citep{GaiaColl2016} started in 2013 and is providing parallaxes and proper motions of $100$ million stars with high spatial resolution, $\sim \mu$as. Until now, {\it Gaia} Data Release 2 (DR2) has been released, and the next date release (DR3), which provides information of binaries, is planned in 2021
\footnote{\url{https://www.cosmos.esa.int/web/gaia/release}}. 
It may include BH binaries with long periods, i.e., days to years.
Moreover, {\it Gaia} can observe black holes-compact objects (such as neutron stars, white dwarfs and brown dwarfs) binaries with a good precision \citep{Andrewsetal2019}.

Some previous researches have estimated the number of black hole-main sequence star binaries (hereafter, BH-MS binaries) detectable with {\it Gaia} \citep{MashianLoeb2017,Breivik2017,Yamaguchi2018,Yalinewich2018,Kinugawa2018,Shao2019}. All of them were considered to be isolated binaries, which were formed as tight binaries and did not experience any dynamical interactions with other stars in the MW disk. \cite{MashianLoeb2017} estimated $2 \times 10^5$ binaries can be detected in $5 \sigma$ sensitivity. \cite{Breivik2017} added the effect of BH natal kick on the number of detectable BH-MS binaries. They found that the number of detectable BH-MS binaries depends on models of BH natal kick by a factor of 3 to 4, and the detection with {\it Gaia} was estimated to be $3,800$--$12,000$, which is one order of magnitude smaller than \cite{MashianLoeb2017}.
Additionally,
  \cite{Breiviketal2019} showed that the observation of such compact objects-giant star binaries will be a good test for wind accretion models.
\cite{Yamaguchi2018} demonstrated various kinds of BH mass distributions and also considered interstellar extinction, which has not been included in the previous work. They showed that $200$--$1,000$ BH-MS binaries can be detected, and that the value varies with the BH mass distribution models.
\cite{Yalinewich2018} predicted {\it Gaia} can observe dozens of BH-luminous companion binaries. They found the companion mass distribution is double-peaked below and above $\sim 10M_\odot$. It was different from \cite{Breivik2017}, which obtained a single-peak distribution in lower mass region ($\lesssim 10M_\odot$). It reflected the difference of a constraint on orbital periods.
\cite{Kinugawa2018} investigated the metallicity dependence of the detectability of BH-MS binaries. They revealed $\sim 200$ binaries with solar metallicity and $\sim 400$ binaries with 0.1 solar metallicity can be detected. The number of detectable binaries is comparable to \cite{Yamaguchi2018}. 
\cite{Shao2019} considered some population synthesis models in terms of BH mass distribution and natal kicks. They showed several hundreds of binaries can be detected.

These previous studies have considered only isolated binaries in the MW disk (field). However, BH-MS binaries are also formed in stellar
clusters through dynamical interactions. 
The aim of this paper is estimating the number of BH-MS binaries formed in open clusters detectable with {\it Gaia}. We also show the distribution of binary parameters such as BH and MS masses, orbital period, and eccentricity and the difference from those of isolated binaries.

This paper consists of the following three parts. In section \ref{sec:method}, 
we introduce our method to obtain observable BH-MS binaries formed 
in open clusters. In section \ref{sec:result}, we show the results 
of {\it N}-body simulation. Finally, we indicate the features of the 
binaries which may reflect the difference in the formation processes in section \ref{sec:discussion}.

\section{Method}
\label{sec:method}
We use the BH-MS binary distribution obtained from $N$-body simulations of open clusters performed in \cite{Kumamoto2020}. 
In this section, we first describe the $N$-body simulation methods used in \cite{Kumamoto2020} (section \ref{sec:Nbodysim}) and the initial condition of the open cluster models (section \ref{sec:IniCon}). 
Finally, we outline how we estimate the number of observable BH-MS binaries 
by using the $N$-body simulation results in section \ref{sec:NumEst}.

\subsection{{\it N}-body Simulation}
\label{sec:Nbodysim}
We use {\sf NBODY6++GPU} \citep{NBODY6++GPU} for the $N$-body simulations.
{\sf NBODY6++GPU} is a direct $N$-body simulation code based on {\sf NBODY6} and {\sf NBODY6-GPU} 
\citep[][respectively]{Aarseth2003,NBODY6-GPU}. It employs a fourth-order Hermite integration method \citep{MakinoAarseth1992} with KS regularization scheme \citep{KustaanheimoStiefel1965,MikkolaAarseth1993}. 
Single and binary stellar evolution models are also included 
\citep[][respectively]{Hurley2000,Hurley2002}. We adopt stellar wind mass loss 
model of \cite{Belczynski2010} and supernova model of \cite{Belczynskietal2002}. 
We switch off BH natal kicks caused by asymmetric supernova explosion for simplicity. We discuss the possible effect of BH natal kicks in Section \ref{sec:detectability}. 

As described above, we choose a supernova model of \cite{Belczynskietal2002}. However, there are newer supernova models, such as \cite{Fryeretal2012}, which can reproduce a BH mass gap between $2$ and $5M_\odot$ \citep{Ozeletal2010,Farretal2011}. In addition, some recent studies showed that Fryers' models do not reflect the importance of the core structure in determining how supernova explosions occur \citep{Uglianoetal2012,PejchaThompson2015,Sukhboldetal2016,Warrenetal2019}. Although our model includes BHs in the mass gap, our choice should have little effect because the most observable BH-MSs have BHs with $\sim 10M_\odot$ as seen below.

\subsection{Initial Condition of Open Clusters}
\label{sec:IniCon}
We adopt an initial mass of our open cluster model ($M_{\mathrm{ini}}$) as $2.5 \times 10^3M_{\odot}$ and the Plummer profile \citep{Plummer1911} for the phase space distribution. We set the initial half-mass density ($\rho_{\mathrm{hm}} \equiv 3M_{\mathrm{ini}}/8 \pi r_{\mathrm{hm}}^3$), where $r_{\mathrm{hm}}$ is the half-mass radius, to be $10^4 M_{\odot} \mathrm{pc}^{-3}$. 
The mass of each particle is randomly assigned following Kroupa initial mass function \citep{KroupaIMF2001}. 
We set the minimum and maximum masses as $0.08M_{\odot}$ and $150M_{\odot}$, respectively. 
Therefore, the average stellar mass ($\langle m\rangle$) is  $0.586M_{\odot}$, and the initial 
number of particles ($N_{\mathrm{ini}}$) is $M_{\mathrm{ini}}/\langle m\rangle = 4266$. 
For the metallicity $Z$, we adopt solar metallicity, $Z = 0.02$. We do not include 
primordial binaries and an external tidal field. 
We perform $1000$ runs with different realizations.

One might think that our cluster model has a mass density much higher than those of currently observed open clusters \citep{PortegiezZwartetal2010}. However, we note that the ages of observed open 
clusters are more than a few Myrs. Even if they initially had a mass density of $\sim 10^4M_\odot\mbox{pc}^{-3}$, their density should have decreased due to gas expulsions, supernovae and dynamical evolution \citep{FujiiPortegiesZwart2016} \citep[see][ for more details]{Kumamoto2020}.

We include no primordial binary as described above. If we included primordial binaries, we would get more detectable BH-MSs than seen below. However, most of them would not be affected by dynamical interactions with other stars. Such binaries evolve similarly with BH-MSs formed in the Galactic field. Although our choice may underestimate the number of detectable BH-MSs originating from open clusters, we do not miss detectable BH-MSs strongly which experienced dynamical interactions in open clusters. In this paper, we focus on these BH-MSs.

\subsection{Number Estimation}
\label{sec:NumEst}
In this subsection, we summarize how we estimate the number of BH-MS binaries observable with $Gaia$ by using the results of our $N$-body simulations. 
We follow \cite{Yamaguchi2018}.

\subsubsection{Distribution in the MW}
\label{sec:distribution}
We distribute BH-MS binaries escaping from open clusters in the MW. 
The number distribution of binaries at a position ($\bm{x}$) is written as a function of BH mass ($m_{\rm BH}$), MS mass ($m_{\rm MS}$), and orbital period ($P$):
\begin{equation}
    n(m_{\mathrm{BH}},m_{\mathrm{MS}}, P, \bm{x}) = 
    \tilde{N}(m_{\mathrm{BH}},m_{\mathrm{MS}}, P) \times t_{\mathrm{MS}} \times \dot{\rho}(\bm{x}),
\end{equation}
where $\tilde{N}(m_{\mathrm{BH}},m_{\mathrm{MS}}, P)$ is defined as
\begin{equation}
    \tilde{N}(m_{\mathrm{BH}},m_{\mathrm{MS}}, P) = 
    N(m_{\mathrm{BH}},m_{\mathrm{MS}}, P) / 1000M_{\mathrm{ini}},
\end{equation}
where $N(m_{\mathrm{BH}},m_{\mathrm{MS}}, P)$ is the number of binaries escaping from all the open clusters in our simulations, $t_{\mathrm{MS}}$ is 
the lifetime of MS and $\dot{\rho}(\bm{x})$ is a formation rate density of stars in open clusters in the MW.
We count only escaping BH-MS binaries, since BH-MS binaries in open clusters are hardly observed. 

We assume that local star-formation rate density is proportional to the local stellar density, and star formation rate ($R_{\rm SF}$) is the same everywhere in the MW disk. Then, 
$\dot{\rho}(\bm{x})$ is given by
\begin{equation}
    \dot{\rho}(\bm{x}) = f_{\mathrm{cl}} \times R_{\mathrm{SF}} \times n_{\mathrm{MW}}(\bm{x}),
\end{equation}
where $f_{\rm cl}$ is the fraction of stars formed in open clusters and $n_{\mathrm{MW}}(\bm{x})$ is the stellar number density distribution at a position ($\bm{x}$) in the MW. 

We consider only the MW disk because the MW bulge cannot be observed 
due to interstellar extinction. In the MW disk, stars exponentially 
distribute from the center of the galaxy \citep{deVaucouleurs1959,Kormendy1977a} 
and perpendicular to the galaxy plane \citep{BahcallSoneira1980}. 
The stellar number density distribution $n_{\mathrm{MW}}(\bm{x})$ 
is expressed as
\begin{equation}
    n_{\mathrm{MW}}(\bm{x}(r, z)) = n_0 \exp \left ( -{{r - r_0}\over{r_h}} - {{z}\over{h_z}} \right ).
\end{equation}
We choose the normalization factor ($n_0$) to satisfy the following equation, 
\begin{equation}
    4 \pi \int^{r_{\mathrm{max}}}_0 rdr \int^{z_{\mathrm{max}}}_0 dz n_{\mathrm{MW}}(\bm{x}) = 1,
\end{equation}
where $r_0$ is a distance from the center of the galaxy to the sun
(8.5~kpc), and $r_h$ and $h_z$ are scale heights to the $r$- and
$z$-directions (3.5~kpc and 250~pc), respectively. We set
$r_{\mathrm{max}} = 30$~kpc and $z_{\mathrm{max}} = 1$~kpc 
\citep[see a review for][]{BlandHawthornGerhard2016}.

For $R_{\rm SF}$, we adopt a total star formation rate in the entire MW disk ($3.5 M_{\odot}\; \mathrm{yr}^{-1}$) \citep{OShaughnessyetal2006b}.
We assume that clusters have the same total mass and density (single population) and that all clusters form binaries following the same binary distributions. We also assume that a certain fraction ($f_{\rm cl}$) of stars are formed in open clusters and that $f_{\rm cl}$ is a constant. We estimate $f_{\mathrm{cl}}$ based on observed star formation rate in the MW and that in open clusters.
According to an observational relationship between the molecular surface density and star formation rate per unit area \citep{Bigieletal2008}, the star formation rate per unit area in the solar neighborhood is estimated as $1.5$ -- $5.1 \times 10^{-3} M_\odot \; \mathrm{yr}^{-1} \; \mathrm{kpc}^{-2}$ assuming the molecular surface density in the solar neighborhood of $3.1 M_\odot \; \mathrm{yr}^{-1}$ \citep{GuestenMezger1982}. 
We also estimate the star formation rate per unit area but only for stars formed open clusters by calculating the total stellar mass of open clusters younger than 100\,Myr within 1\,kpc from the Sun. By using a catalog of open clusters \citep{Piskunov2007}, we find that $6.6 \times 10^{-4}M_{\odot} \; \mathrm{yr}^{-1}$ for stars in open clusters.
Therefore, we estimate $f_{\mathrm{cl}}\sim 0.1$.

The total number of BH-MS binaries detectable with $Gaia$ ($N_{\rm det}$) is obtained as
\begin{eqnarray}
    &&N_{\rm det} = \int dm_{\mathrm{BH}} \int dm_{\mathrm{MS}} \int dP \nonumber \\
    && \;\;\;\;\;\;\;\;\;\;\;\;\;\;\;\;\;\;\;\;\;\;\;\;\;\; \times \int_{|\bm{x}-\bm{x}_0|<D_{\max}} d^3 \bm{x} n(m_{\mathrm{BH}},m_{\mathrm{MS}},P,\bm{x}),
    \label{number_estimation}
\end{eqnarray}
where $\bm{x}_0$ is the position of the Sun and $D_{\max}$ is 
the maximum distance dependent on the parameters of the binaries. 
From the next section, we evaluate $D_{\max}$ by considering some conditions: interstellar extinction and observational constraints.

\subsubsection{Interstellar Extinction}
\label{sec:stellarext}
We calculate the maximum distance at which the MS of a binary 
is luminous enough to be observed with $Gaia$. Here, we call it $D_{\rm MS}$. Considering interstellar extinction, $D_{\mathrm{MS}}$ depends on MS mass ($m_{\mathrm{MS}}$), and  
it satisfies the equation below,
\begin{equation}
    m_{\mathrm{V}}(m_{\mathrm{MS}},D_{\mathrm{MS}}) = m_{\mathrm{v,lim}},
\end{equation}
where $m_{\mathrm{v,lim}}$ is the maximum apparent magnitude
observable with $Gaia$. We adopt $m_{\rm v, lim}=20$ \citep{GaiaColl2016}. 
Following \cite{Yamaguchi2018}, we adopt V band instead of the {\it Gaia} band because the color $V-I$ is less than 1 \citep{Jordietal2010}. We use the empirical relationship 
between MS mass ($m_\mathrm{MS}$) and its absolute magnitude in V band 
($M_\mathrm{V}$) \citep{Smith1983} is
\begin{equation}
    m_{\mathrm{MS}} = \left \{ \begin{array}{ll}
            10^{-0.1(M_\mathrm{V} - 4.8)} & (M_\mathrm{V} < 8.5) \\
            1.9 \times 10^{-0.17(M_\mathrm{V} - 4.8)} & (M_\mathrm{V} > 8.5).
        \end{array} \right .
\end{equation}

The apparent magnitude ($m_{\mathrm{V}}$) is calculated by using
\begin{equation}
    M_\mathrm{V} = m_\mathrm{v} - 5(2 + \log_{10}D_{\mathrm{kpc}}) - A_\mathrm{V}(D_{\mathrm{kpc}}),
\label{ext}
\end{equation}
where $D_{\mathrm{kpc}}$ is the distance to the binary in kpc and
$A_\mathrm{V}$ is interstellar extinction. Since the average
extinction in the MW is $\sim 1$\,mag per 1\,kpc in V band
\citep{Spitzer1978,Shafter2017}, we adopt $A_\mathrm{V} \sim
D_{\mathrm{kpc}}$. Therefore, we obtain $D_{\mathrm{MS}}$ 
satisfying the equation:
\begin{equation}
  M_{\rm V} + 5(2 + \log_{10}D_{\rm MS}) + D_{\rm MS} = 20,
\end{equation}
where $D_{\rm MS}$ is in kpc.

\subsubsection{Observational Constraints on Detection of BHs}
\label{sec:obsconst}
We identify a binary with a MS and an unseen object by observing
the motion of the MS. In order to determine the unseen object as a BH, 
the lower mass limit of the unseen object should be heavier than
$3M_\odot$ \citep{KalogeraBaym1996}. In other words, the estimated
mass of the unseen object should satisfy an equation:
\begin{equation}
    m_{\mathrm{BH}} - n \sigma_{\mathrm{BH}} > 3M_{\odot},
\label{BHconst}
\end{equation}
where $m_{\mathrm{BH}}$ is the mass of an unseen object and
$\sigma_{\mathrm{BH}}$ is its standard error. We set $n=1$
following \cite{Yamaguchi2018}.

The observables of a binary are the MS mass
($m_{\rm MS}$), the binary orbital period ($P$), the angular semi-major axis ($a_*$), and the distance to the binary ($D$). 
Their relationship is expressed as 
\begin{equation}
    {{(m_{\mathrm{MS}} + m_{\mathrm{BH}})^2}\over{m_{\mathrm{BH}}^3}} = {{G}\over{4 \pi^2}} {{P^2}\over{(a_* D)^3}},
\label{binary}
\end{equation}
where $G$ is the gravitational constant. From this equation, we relate the standard error of BH mass ($\sigma_{\rm BH}$) to
standard errors of other binary parameters. We can derive the relationship of the standard error of each parameter:
\begin{eqnarray}
     &&\left( {{\sigma_{\mathrm{BH}}}\over{m_{\mathrm{BH}}}} \right )^2 = \left( {{3}\over{2}} - {{m_{\mathrm{BH}}}\over{m_{\mathrm{BH}}+m_{\mathrm{MS}}}} \right )^{-2} \times \nonumber \\
     && \left[ \left( {{m_{\mathrm{MS}}}\over{m_{\mathrm{BH}}}+m_{\mathrm{MS}}} \right )^2 {{\sigma_{\mathrm{MS}}^2}\over{m_{\mathrm{MS}}^2}} + {{\sigma_{P}^2}\over{P^2}} + {{9}\over{4}} \left( {{\sigma_{a*}^2}\over{a_*^2}} + {{\sigma_{D}^2}\over{D^2}} \right) \right].
     \label{mbh}
\end{eqnarray}
where $\sigma_{\rm MS}$ is a standard error of the 
MS mass, $\sigma_{P}$ is that of the binary orbital period, $\sigma_{a*}$ is that of the angular semi-major axis, and $\sigma_{D}$ is that of the distance. 
Here, we assume that each standard error is sufficiently smaller than 
each observable, and ignore the correlation among the observables.

If we assume that each error should be smaller than 10\% of the value of each variable for detection, that is, 
\begin{equation}
    {{\sigma_{\mathrm{MS}}}\over{m_{\mathrm{MS}}}} < 0.1, 
    {{\sigma_{P}}\over{P}} < 0.1, {{\sigma_{a*}}\over{a_*}} < 0.1,
    \;\; \mathrm{and } \; \; {{\sigma_{D}}\over{D}} < 0.1,
    \label{const}
\end{equation}
BHs with masses of $\gtrsim 3.75M_{\odot}$ can be detected. 
There are only few BHs lighter than $5M_{\odot}$ \citep{Ozeletal2010}. 
Therefore, we consider the constraints shown in equation (\ref{const}) as sting enough for detection of all BH-MS binaries.

In particular, the first and second conditions in equation~(\ref{const}) 
are easily achieved. According to \cite{Tetzlaffetal2011}, a typical 
standard error of a stellar mass estimated from its spectral and luminosity 
is smaller than $10 \%$. That of an orbital period can be limited to $\lesssim 10 \%$ 
if the observed period is shorter than two-thirds of the observational time 
\citep{ESA1997}. For {\it Gaia}, it is 5 years and we adopt 3 years 
($\sim 5 \times 2/3$) to the maximum period. 
We also set the minimum period as 1 day. This minimum period may be too short, since {\it Gaia}'s cadence for each object is about several ten days. However, this choice has little effect on the number of observable BH-MSs. BH-MSs with periods shorter than several ten days cannot be observed due to the lower limit of observational constraints from orbital separations described below.

For the rest conditions, we can obtain additional constraints on
$D_{\mathrm{max}}$. Since a parallax $p$ in arcsec is equal to $1/D$, the distance to the binary $D$ in pc, we
can derive the following equation:
\begin{equation}
  \frac{\sigma_p}{p} \sim \frac{\sigma_D}{D} < 0.1, \label{eq:ParallaxAndDistance}
\end{equation}
where $\sigma_{p}$ is a standard error of a
parallax. \cite{GaiaColl2016} have expressed that in G band $\sigma_{p}$ as
\begin{equation}
    \sigma_p = (-1.631 + 680.8z(m_\mathrm{v}) +
    32.73z(m_{\mathrm{v}})^2)^{1/2} [\mu \mathrm{as}], \label{eq:StandardErrorOfParallax}
\end{equation}
where
\begin{equation}
    z(m_{\mathrm{v}}) = 10^{0.4 (\mathrm{max}[12.09,m_{\mathrm{v}}]-15)}.
\end{equation}
Note that we neglect the dependence on the $(V-I)$ color. Substituting
equation~(\ref{eq:StandardErrorOfParallax}) into
equation~(\ref{eq:ParallaxAndDistance}), we obtain the condition of a
distance as
\begin{equation}
    (-1.631 + 680.8z(m_\mathrm{v}) + 32.73z(m_{\mathrm{v}})^2)^{1/2} <
  {{10^2}\over{D_{\mathrm{kpc}}}}.
  \label{Breivik_only}
\end{equation}

We note the right-hand side of the above equation ($0.1 p [\mu \mathrm{as}]$) is obtained as follows:
\begin{eqnarray}
  p [\mathrm{as}] &=& 1/D_{\rm pc} = 10^{-3}/D_{\rm kpc} \nonumber \\
  p [\mu \mathrm{as}] &=& 10^6 \times 10^{-3}/D_{\rm kpc} =  10^3/D_{\rm kpc} \nonumber \\
  0.1p [\mu \mathrm{as}] &=& 10^2/D_{\rm kpc},
\end{eqnarray}
where $D_{\rm pc}$ and $D_{\rm kpc}$ are a distance to a binary in the units of pc and kpc, respectively.
Hereafter, the maximum distance satisfying this equation is called $D_{p}$.

Finally, the constraint from a semi-major axis set another constraint on $D_{\max}$. 
Since the semi-major axis of each binary is close to the orbital radius 
on celestial sphere, the standard error of semi-major axis ($\sigma_{a*}$) 
is $\sim \sigma_{p}$. Therefore, the constraint from the semi-major axis of 
a binary which can be observed with {\it Gaia} is
\begin{equation}
    a > 10 {{m_{\mathrm{BH}} + m_{\mathrm{MS}}}\over{m_{\mathrm{BH}}}} D \sigma_p,
    \label{Yamaguchi_only}
\end{equation}
and we obtain the maximum distance satisfying the equation above and assign it to $D_a$.

Therefore, by substituting $D_{\mathrm{MS}}, D_{p},$ and $ D_a$ to
$D_{\mathrm{max}}$, we can include each constraint. We set the maximum
distance considering no effects as 10~kpc.

\section{Results}
\label{sec:result}

\subsection{Properties of Escaping BH-MS Binaries}

\begin{figure}[tp]
\begin{center}
    \includegraphics[width=80mm]{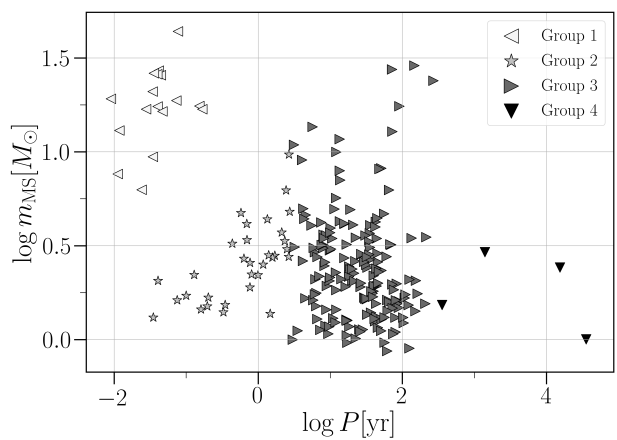}
    \caption{MS mass $m_{\mathrm{MS}}$ and orbital period $P$ distribution of the binaries escaping from all the open clusters.}
    \label{all0}    
\end{center}
\end{figure}

We demonstrated $N$-body simulation of open cluster models which include $4266$ initial stars for $1000$ different realizations. We did not include any primordial binaries.
According to the results of {\it N}-body simulations, we investigate the parameters of BH-MS binaries escaping from all the open clusters.

We obtain 225 BH-MS binaries in total from all the open clusters. Figure \ref{all0} shows MS mass and orbital period distribution of them. We divide them into four groups.
Since we assume that {\it Gaia} can observe binaries whose orbital periods are $\sim 1$ day to 3 years, we separate the binaries mainly based on their orbital period. 
We call a group of binaries with heavier MS
masses ($m_{\mathrm{MS}} \gtrsim 5.6M_{\odot}$) and orbital periods shorter than
3 years as Group 1. 
We categorize binaries with orbital periods shorter than $\sim 3$\,years but with lighter MS masses ($m_{\mathrm{MS}} \sim 1.8-5.6M_{\odot}$) as Group 2.
Most of the other binaries have orbital periods between 3 years and 270 years. We call them Group 3. We also find only a few binaries with extremely long orbital 
periods ($P \gtrsim$300 years) and classify them as Group 4.

We find that Group 1 binaries are formed through three-body encounters and binary-single interactions when BH progenitors were still MSs.
They experience common envelope phases, and one of the binary MSs evolve to BHs.
Until they escape from open clusters, they have not closely interact with any other stars.
Thus, Group 1 binaries are a kind of isolated binaries after they were formed.

\begin{figure}[tp]
\begin{center}
    \includegraphics[width=80mm]{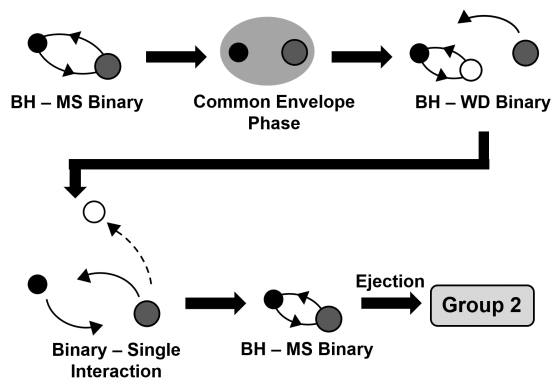}
    \caption{The scheme of the formation process of Group 2 binaries.}
    \label{group2_formation}    
\end{center}
\end{figure}

Most of Group 2 binaries are formed as follows. 
Single massive stars evolve to BHs and capture MSs to form binaries.
They experience common envelope phase, and evolve to BH-compact object such as white dwarf (WD) binaries. Figure \ref{group2_formation} is the schematic figure of this formation process.
The rest of Group 2 binaries experience common envelope phases, while they are 
MS-MS binaries and evolve to BH-MS binaries. Finally, all of Group 2 binaries exchange one of their members with MSs through binary-single interactions and escape from open clusters.

The mass gap between Groups 1 and 2 is a trade-off region. A MS in such mass region is neither heavy enough to form a binary nor light
enough to be weighted by the initial mass function (heavier stars are less likely to be formed).

Group 3 binaries were formed by three-body encounters and binary-single encounters
including BHs and MSs. The BHs evolved from MSs 
in single stellar evolution and captured MSs. Group 3 binaries have orbital periods 
characterized by the escape velocities of open clusters.

Group 4 binaries are first formed via three-body encounters. Three of them were still MS-MS binaries when they were formed, and the companions of them evolved to BHs. The BH of the rest were formed from single stars, and then they formed binaries with MS stars due to three-body encounters. All of Group 4 binaries exchanged their members with other MSs via binary-single interactions before they escaped from open clusters. They may have escaped from open clusters at the moment the clusters evaporate.

Binaries in all the Groups have BH masses in a range from
$3M_{\odot}$ to $20M_{\odot}$. The mass distribution of BHs are very similar among the four Groups.

\subsection{BH-MS Binaries Observable with {\it Gaia}}
\label{sec:detectability}

\begin{table}
      \tbl{The total number of BH-MS binaries observable with {\it Gaia}
      ($N_{\rm det}$) for each model. "10\,kpc model": the number of
      binaries within 10\,kpc from the Sun, "Breivik model": the number 
      with the observational constraint from parallaxes in $3 \sigma $
      confidence level. "Yamaguchi model": the number with interstellar extinction and the observational constraints from parallaxes and orbital separations. \label{number_group}}{
    \begin{tabular}{lc|cc} \hline
           & $N_{\rm det}$ & Group 1 & Group 2 \\ \hline
          $10$\,kpc model & $2.0 \times 10^3$ & $14$ & $2.0 \times 10^3$ \\
          Breivik model & $ 7.1 \times 10^2$ & $14 $ & $6.9 \times 10^2 $ \\
	      Yamaguchi model & $8.9$ & $5.6 \times 10^{-4}$ & $8.9$ \\ \hline
    \end{tabular}}
\end{table}

\begin{figure}[tp]
    \begin{center}
    \includegraphics[width=80mm]{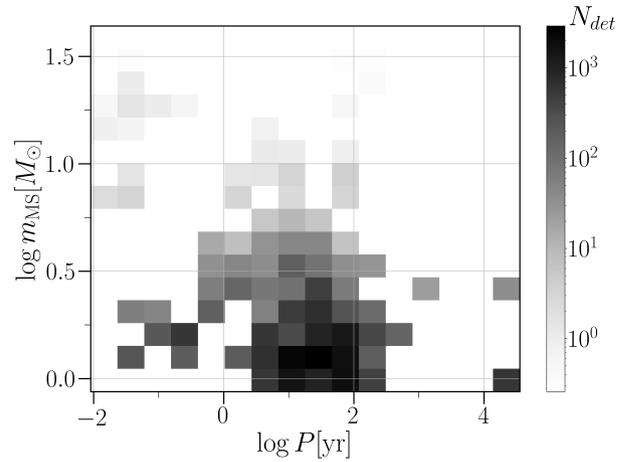}
    \end{center}
      \caption{Two-dimensional histogram for the numbers of all the
    escaping binaries from open clusters within 10~kpc from the
    Sun. The gray scale shows the number of BH-MS binaries supposed to exist within 10~kpc from the Sun. 
    }
    \label{det_noext}
\end{figure}

\begin{figure}[tp]
    \begin{center}
    \includegraphics[width=80mm]{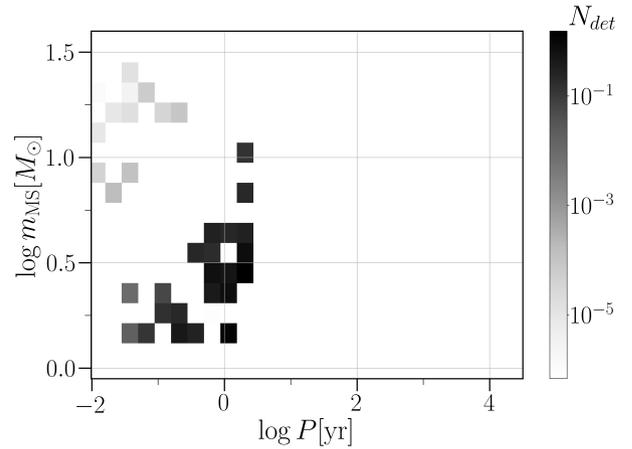}
    \end{center}
      \caption{The same as Figure~\ref{det_noext} but for the numbers of
    binaries detectable with {\it Gaia}. We consider both of
    interstellar extinction, parallax limit, and orbital separation
    limit. (corresponding to the Yamaguchi model)}
    \label{det_wallerr}
\end{figure}

In this section, we estimate the number of BH-MS binaries for the following constraints. We first count the number of binaries within 10\,kpc from the Sun by taking the lifetime of the MS into account. Then, we also add the effects of interstellar extinction and observational constraints (see sections \ref{sec:stellarext} and \ref{sec:obsconst} for more details). Note that we consider only Group 1 and 2 binaries as detectable with {\it Gaia} since we adopt a constraint from orbital periods (see section \ref{sec:obsconst}).

Table \ref{number_group} shows the total expected number of BH-MS binaries within 10~kpc from the Sun and of detectable binaries with {\it Gaia} ($N_{\rm det}$). We also show the respective numbers of Groups 1 and 2 binaries under different restrictions. 
``10\,kpc model'' indicates all the binaries within 10~kpc from the Sun (without interstellar extinction and any observational constraints), where $D_{\max}=10$~kpc in equation (\ref{number_estimation}). 
``Breivik model'' denotes binaries detectable in $3\sigma$ confidence level when we take into account only the observational constraint from parallax (described in equation (\ref{Breivik_only}) and relax the constraint to $3 \sigma$), where $D_{\max}= \mathrm{min}(10,D_p)$~kpc in equation (\ref{number_estimation}).
``Yamaguchi model'' shows binaries observable, when we take into account interstellar extinction (section \ref{sec:stellarext}) and all the observational constraints (see equations (\ref{Breivik_only}) and (\ref{Yamaguchi_only}) of section \ref{sec:obsconst}) where $D_{\max}=\mathrm{min}(10,D_{\mathrm{MS}},D_p, D_a)$~kpc in equation (\ref{number_estimation}). 

The total number of BH-MS binaries formed in open clusters within $10$\,kpc from the Sun is $\sim 2 \times 10^3$. Even if we take into account the most strict restrictions related to interstellar extinction and the observational constraints from parallaxes and orbital separations, we estimate that {\it Gaia} can detect $\sim 10$ BH binaries originating from open clusters (see the Yamaguchi model). If we relax the restrictions, we predict that {\it Gaia} will be able to observe $\sim 7 \times 10^2$ BH binaries (see the Breivik model).

We find that Group 2 is dominant in all the types of the restrictions. Since binaries in Group 2 have less massive MSs and longer orbital periods, the MSs live longer and the observational constraint from orbital separations does not suppress the number of Group 2 binaries very much. 

For the Breivik model, the number of Group 1 binaries does not decrease compared to that for the 10~kpc model. The observational constraint (from parallaxes) for the Breivik model is not as strict as those of the Yamaguchi model. The number of Group 2 binaries becomes one-third of that of the 10~kpc model. This is because MSs of Group 2 binaries are less massive and fainter.
On the other hand, the number of Group 1 and 2 binaries in the Yamaguchi model decrease drastically. In particular, the number of Group 1 becomes $10^{-5}$ of that of the Breivik model. Since Group 1 has tighter binaries than Group 2, most of the binaries in Group 1 cannot be observed due to the lower limit of the observational constraint from orbital separations.

Figure~\ref{det_noext} shows two-dimensional histogram of BH-MS binaries within 10\,kpc from the Sun (the 10\,kpc model). 
Here, we assign the distribution in Figure \ref{all0} to each period-mass bin and calculate the number of binaries in each bin. Here, the lifetime of MS of the binaries is considered, but interstellar extinction and any observational constraints have not been included yet. 
In this plot, the number of binaries with less massive MSs is larger than that of binaries with more massive MSs despite that the numbers are very similar when they are formed in clusters (see Figure~\ref{all0}). This is
because the lifetime of low-mass MS is longer than that of massive MS.

In Figure~\ref{det_wallerr}, we present the two-dimensional histogram for the number of binaries detectable with {\it Gaia} for the Yamaguchi model. Compared with Figure~\ref{det_noext}, a significant fraction of binaries in Group 1 and 2 become difficult to observe.
In particular, as seen in Table \ref{number_group}, Group 1 binaries are hardly observed because of the lower limit of the observational constraints from orbital separations.
Therefore, we expect that BH-MS binaries formed in open clusters should be found in the MS mass-period region of Group 2. 

\begin{figure}[tp]
    \begin{center}
    \includegraphics[width=80mm]{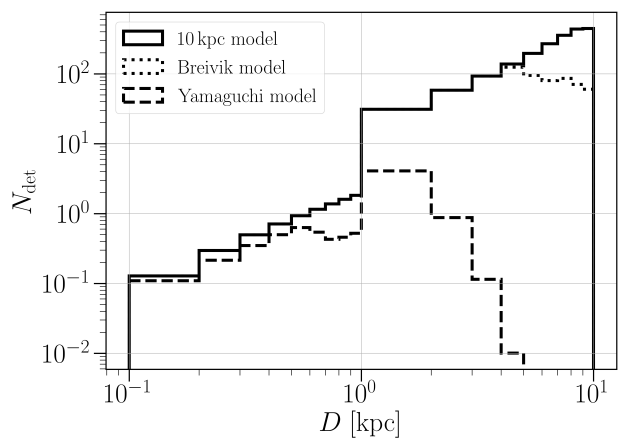}
    \caption{The number distribution of BH-MS binaries as a function of the 
    distance from the Sun. Solid histogram shows the number of binaries within 
    10~kpc from the Sun. Dotted one 
    describes the number considering the observational constraint from parallaxes 
    in $3 \sigma $ confidence level (the Breivik model), and dashed one shows the number with interstellar extinction and the observational constraints from parallaxes and orbital separations (the Yamaguchi model).}
    \label{kpc}
    \end{center}
\end{figure}

In the following, we describe binary parameter distributions of observable BH-MS binaries.
We present the distance distribution of binaries from the Sun in Figure \ref{kpc}. Without observational constraints and interstellar extinction, the number of binaries monotonically increases as we include distant binaries as seen for the 10\,kpc model, simply because the integrated volume also increases. For the Breivik model, we find that most of binaries are still observable. For the Yamaguchi model, however, the number of observable BH-MS binaries dramatically drops over a few kpc from the Sun. In addition, no BH binaries farther than $\sim 6$\,kpc are observable due to the lower limit of the observational constraint from orbital separations. Since natal kicks caused by fallback materials (hereafter, FB kicks) might increase orbital separations, the observable distance would get further.

\begin{figure}[hbtp]
    \begin{center}
    \includegraphics[width=80mm]{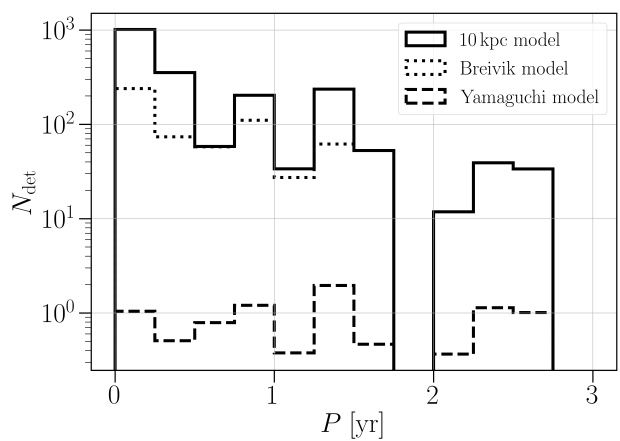}
    \end{center}
    \caption{Period distribution of BH-MS binaries. Solid histogram represents the $10\;$kpc model, dotted one shows the Breivik model, and dashed one describes the Yamaguchi model.}
    \label{peri}
\end{figure}

We can confirm that the decrease of short-period binaries because of the lower limit of the observational constraints from orbital separations.
In Figure \ref{peri}, we show the orbital period distribution of each model. For
the Yamaguchi model, the number of binaries with short periods ($\lesssim 1.5$ yr) decrease by almost three orders of magnitude, while the number of binaries with long periods ($\gtrsim 1.5$ yr) does only by two orders of magnitude.
The gap between $1.75 - 2$ years just reflects a small number statistics.

\begin{figure}[tp]
    \begin{center}
    \includegraphics[width=80mm]{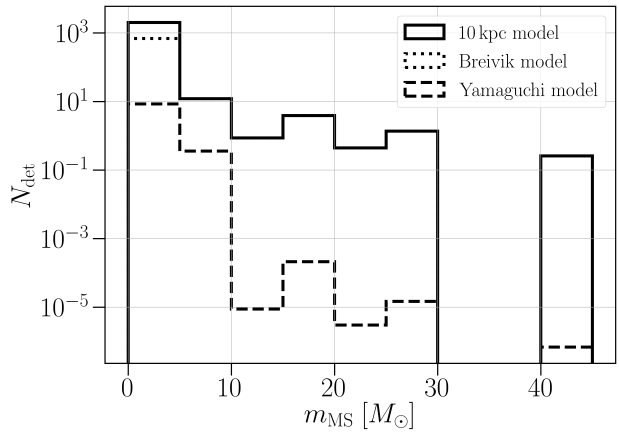}
    \end{center}
    \caption{MS mass distribution of BH-MS binaries. Solid histogram represents the $10\;$kpc model, dotted one shows the Breivik model, and dashed one describes the Yamaguchi model.}
    \label{ms}
\end{figure}

We also investigate the distribution of masses of binaries originating from open clusters. 
Figure \ref{ms} presents the MS mass distribution for each observation model. Without any observational constraints (the 10\,kpc model), low-mass MS ($\lesssim 5M_{\odot}$) is the most populous because of the initial mass function and their long lifetimes, and a few tens of BH-MS binaries with a high-mass MS ($\gtrsim 5M_\odot$) are predicted. For the Breivik model, the distribution does not change except for the least massive stars.
For the Yamaguchi model, only $\sim 10$ BH-MS binaries with a low-mass MS are detectable with {\it Gaia}. In addition, high-mass MS-BH binaries would not be detected. These high-mass MS-BH binaries belong to Group 1 and therefore have a short period. They are hardly observed due to 
the lower limit of 
the observational constraint from the orbital separation as described above.

\begin{figure}[tp]
    \begin{center}
    \includegraphics[width=80mm]{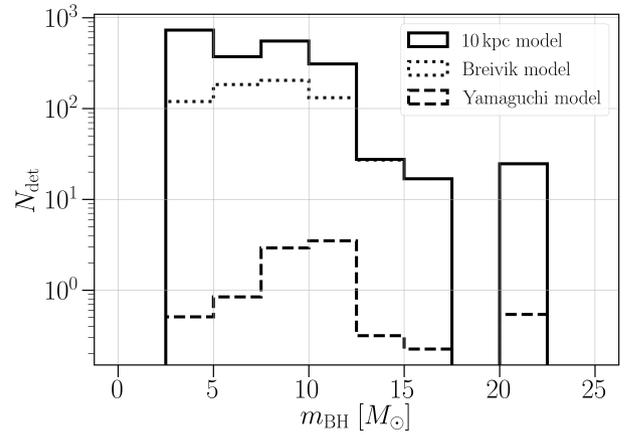}
    \end{center}
    \caption{BH mass distribution of BH-MS binaries. Solid histogram represents the $10\;$kpc model, dotted one shows the Breivik model, and dashed one describes the Yamaguchi model.}
    \label{bh}
\end{figure}

We also present the BH mass distribution expected to be observed in Figure \ref{bh}. The distribution without observational constraints shows a flat distribution in the range from $\sim 2.5M_\odot$ to $\sim 13M_\odot$. This is similar for the Breivik model. For the Yamaguchi model, on the other hand, the most detectable binaries have BHs with mass of $\sim 10M_\odot$. This can be explained as follows. Fewer binaries with more massive BH than $\sim 10M_\odot$ exist compared with those with less massive BHs (see the 10\,kpc model) because of the initial mass function. On the other hand, BH binaries with more massive BHs can be more easily detected, since they swing around their MS companions more largely. Thus, detectable BH mass distribution using binary observation have a peak at $\sim 10M_\odot$. The lower mass limit of detectable BH mass depends on equation (\ref{BHconst}) in our model. Although we set $n=1$ in equation (\ref{BHconst}), $n$ can be larger. If we adopt $n=2$ or $3$, the lower mass limit are $5M_\odot$ and $7.5M_\odot$, respectively. These are still smaller than the peak in the BH distribution at around $10M_\odot$, and therefore the choice of $n$ does not affect the total detectability very much.
In addition, FB kicks would affect the detectability. When BHs with masses of around $10M_\odot$ in Group 2 binaries are formed, their escaping velocity is $\lesssim 20$\,km/s, smaller than their FB kicks estimated from neutron star natal kick \cite{Hobbsetal2005}. Therefore, the detectability including FB kicks would decrease, but some of them would remain.

\begin{figure}[tp]
    \begin{center}
    \includegraphics[width=80mm]{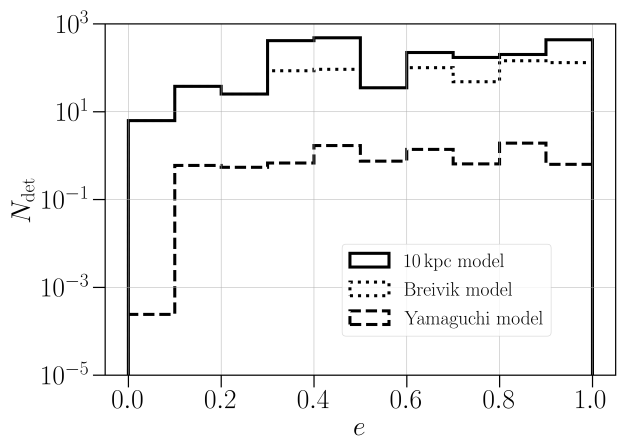}
    \end{center}
    \caption{Eccentricity distribution of BH-MS binaries. Solid histogram represents the $10\;$kpc model, dotted one shows the Breivik model, and dashed one describes the Yamaguchi model.}
    \label{ecce}
\end{figure}

In the eccentricity distribution, we find a characteristic of binaries dynamically formed in open clusters. In Figure \ref{ecce}, we present the eccentricity distribution. In all the observation models, BH-MS binaries tend to have high eccentricities ($e \gtrsim 0.1$). Especially in the Yamaguchi model, BH-MS binaries rarely have nearly zero eccentricities. 
In general, dynamically formed binaries have non-zero eccentricities, and therefore the number of BH-MS binaries with nearly zero eccentricities is initially small. On the other hand, BH-MS binaries which experience common envelope phases have nearly zero eccentricities. However, such binaries tend to be very tight and have too short a period to be observed.
Thus, we predict that BH-MS binaries detectable with {\it Gaia} tend to be eccentric. 

Several studies have shown that binary masses can be estimated even if their phase-coverage orbits are low, $\lesssim 50$ \% \citep[e.g.][]{Lucy2014,ONeiletal2019}. Thus, BH-MSs with periods of longer than $3$ years can be detectable with {\it Gaia}. Moreover, {\it Gaia} mission may have a longer lifetime than expected \footnote{\url{https://www.cosmos.esa.int/web/gaia/release}}. Here, we relax the upper limit of detectable orbital periods.
If we change the upper limit of detectable orbital period to five, ten and twenty years, it is natural that more binaries can be detected because of the contribution of longer orbital period binaries. Companion MSs in such binaries are easier to detect because they moves more largely. The total number of BH-MS binaries are shown in Table \ref{number_periodchange}.

\begin{table}
      \tbl{The total number of BH-MS binaries observable with {\it Gaia}
      ($N_{\rm det}$) for Yamaguchi model with different upper limit of orbital period. "3 yrs" corresponds to the model we discussed in Section \ref{sec:detectability}. \label{number_periodchange}}{
    \begin{tabular}{lcccc} \hline
           & 3 yrs & 5 yrs & 10 yrs & 20 yrs \\ \hline
          $N_{\rm det}$ & 8.9 & 15 & 58 & $1.0 \times 10^2$\\ \hline
    \end{tabular}}
\end{table}

\section{Discussion}
\label{sec:discussion}
Hereafter, we discuss the differences between BH-MS
  binaries formed in the MW galactic field and in open clusters. A possible
  clue can be eccentricities of BH binaries. Figure 4 of
  \cite{Breivik2017} shows that BH binaries formed in a galactic field
  prefer nearly zero eccentricities, while those formed in open
  clusters have high eccentricities at high probability. 
  If a BH binary with a nearly zero eccentricity is
  discovered, its origin must be a galactic field. If a BH
  binary with high eccentricity is found, it could be formed on a
  galactic field or in an open cluster. By comparing the number of BH binaries with
  eccentricities of $\gtrsim 0.1$ in Figure 4 of \cite{Breivik2017}
  and that of our Breivik model in Figure \ref{ecce}, we can find the numbers
  of the former and latter are $\sim 10^3$,
  comparable. Hence, half of BH binaries with high eccentricities
  originate from open clusters.

Another possible clue is the chemical abundance pattern of companion stars of BH binaries. In the case of detectable field-origin BH-MSs, the MSs can be  polluted by stellar wind and (failed) superanova ejecta from the BH progenitors. This is analogy from the chemical anomalies of companions in low-mass X-ray   binaries (LMXBs) with BHs \citep{Israelian1999,Orosz2001,GonzalezHernandez2004,GonzalezHernandez2005,GonzalezHernandez2006, GonzalezHernandez2011}  \citep[see][for a review]{Casares2017}. Note that companions in BH binaries  observable with {\it Gaia} might be less polluted than those in LMXBs,  since the former systems have larger distances than the latter systems. On the other hand, detectable  cluster-origin BH binaries are formed, such that the BH progenitors evolve to BHs, and subsequently they get their final companions. Thus, the BH progenitors do not pollute their final companions. In summary, if a companion star in a BH binary has a normal chemical abundance, the BH binary should be formed in open clusters.

We can see from Figure 2 of \cite{Breivik2017} that
  BH binaries formed in the Galactic field have a period gap around a
  period of $\sim 1.5$ year, if their companion masses are $\lesssim
  5M_\odot$. The reason why there is no period gap at $P \sim 1.5$
  year in Figure 4 of \cite{Breivik2017} is that the period gap is
  bridged by BH binaries with $m_{\rm MS} \gtrsim 5M_\odot$. We
  interpret this period gap as follows. Binaries with orbital periods
of $\lesssim 1.5$ year at the initial time become shorter-period
binaries due to common envelope evolution, while binaries with
$\gtrsim 1.5$ year keep their periods due to the absence of binary
interactions. On the other hand, binaries in Group 2 do not have such
a period gap. This is because they are formed through both of binary and dynamical interactions (see Figure
\ref{group2_formation}). 
It is worth performing follow-up observations of such BH binaries to confirm their origins. The follow-up observations can be, for example, determination of their chemical abundance pattern as described above.

\section{Conclusion}
In this paper, we estimated the number of BH-MS binaries formed in open clusters detectable with {\it Gaia} and investigated their properties.
We used the results of {\it N}-body simulations of open clusters as the properties of binaries produced in them. 
We assumed that local star-formation rate was proportional to the local stellar density and that the star formation rate in the past was constant. We also assumed that the fraction of stars forming open clusters was constant.
By using the results of our simulations and these assumptions, we estimated the parameter distributions of BH-MS binaries formed in open clusters. 

The results of {\it N}-body simulation indicated that the detectable BH-MS binaries consist of two groups. Group 1 binaries have heavier MS mass ($m_{\mathrm{MS}} \gtrsim 5.6M_{\odot}$). They are formed through three-body encounters and binary-single encounters when they are still MS-MS binaries. They experience common envelope phases and then one of binary stars evolve to BHs. Their evolution is similar to that of isolated field binaries. Group 2 binaries have less massive MS masses ($m_{\mathrm{MS}} \sim 1.8-5.6M_{\odot}$) than those of Group 1. Most of Group 2 binaries also experience common envelope phases, but they later exchange MSs before escaping from open clusters.
These two types of BH-MS binaries can be detected with {\it Gaia}. 
We estimated $2.0 \times 10^3$ BH-MS binaries exist within $10\;$kpc from the Sun. Group 2 binaries are dominant because they have less massive MSs and as a result, they have longer lifetimes.

We also considered the effect of interstellar extinction and the observational constraints following \cite{Yamaguchi2018}.
By considering these, we predicted that $\sim 10$ BH-MS binaries can be detected with {\it Gaia} and that they are rarely detected further than 6~kpc from the Sun.
In particular, Group 1 binaries are hardly detected. They are tighter binaries than those in Group 2, and therefore the observational constraint from orbital separations makes them much difficult to be detected. 

We suggested chemical abundance patterns of companion
  MSs may help us to identify the origin of binaries.
Group 2 binaries exchange MSs through binary-single interaction after
common envelope phases. Therefore, companion MSs are supposed to have
normal chemical abundances. Such binaries may not be
formed via isolated binary evolution according to
  analogy from the chemical anomaly of companions in LMXBs with
  BHs. Finally, we found that BH-MS binaries with less massive MSs
  ($m_{\mathrm{MS}} \lesssim 5M_{\odot}$), orbital periods ($P \sim
  1.5\;$year) and higher eccentricities ($e \gtrsim 0.1$) can be
  dominated by cluster-origin BHs. When such BH binaries are
  discovered, it is worth performing follow-up observations to confirm
  their origin. The follow-up observations can be detemination of
  chemical abundance pattern of their companion stars.

\section*{Acknowledgement}
We greatly thank the anonymous referee for useful comments.
We thank N. Kawanaka for fruitful discussion. This work is supported by JSPS 
KAKENHI Grant Number 17H06360, 19H01933, and 19K03907. Numerical calculations in this work was conducted in support of Promotion of Young or Women Researchers, 
Supercomputing Division, Information Technology Center, The University of Tokyo.

\bibliographystyle{apj}
\bibliography{reference}

\end{document}